\documentstyle[preprint,aps,epsf,floats]{revtex}

\begin{document}

\preprint{\tighten \vbox{\hbox{} }}

\title{Remarks on Form Factor Bounds}
\author{Cheng-Wei Chiang\footnote{\tt chengwei@andrew.cmu.edu}}

\address{
Department of Physics,
Carnegie Mellon University,
Pittsburgh, PA 15213}

\maketitle

{\tighten
\begin{abstract}
Improved model independent upper bounds on the weak transition form
factors are derived using inclusive sum rules.  Comparison of the new
bounds with the old ones is made for the form factors $h_{A_1}$ and
$h_V$ in $B \to D^*$ decays.
\end{abstract}
}


\newpage

A set of model independent bounds has been derived to provide a
restriction on the shape of weak transition form factors
\cite{BSUV95,BR9798,BLRW97}.  They have been extensively used to bound
weak decay form factors and the decay spectrum of heavy hadrons
\cite{BR9798,BLRW97,CL99,C99}\footnote{See, however,
\cite{BGL96,CLN98}for model independent parametrizations of the form
factors}.  Here We provide a more stringent upper bound without any
further assumptions.  This upper bound differs from the one derived
previously at order $1/m_Q^2$ or $\alpha_s/m_Q$.  Though this is only
a small improvement, it is worth doing because it can give a tighter
bound from above if one includes higher order corrections.


The bounds are derived from sum rules that relate the inclusive decay rate,
calculated using the operator product expansion (OPE) \cite{W69,M94} and
perturbative QCD, to the sum of exclusive decay rates.  To be
complete, we will derive both the upper and lower bounds, though the
lower bound is the same as the previous one.

Without loss of generality, we take for example the decay of a $B$
meson into an $H$ meson, with the underlying quark process $b \to f$,
where $f$ could be either a heavy or light quark.  First, consider the
time ordered product of two weak transition currents taken between two
$B$ mesons in momentum space,
\begin{eqnarray}
\label{structurefn}
T^{\mu\nu}& = & - \frac{i}{2M_{B}} \int d^4x \, e^{-iq \cdot x} 
   \left< B(v) \right| T(J^{\mu\dagger}(x)J^{\nu}(0)) \left| B(v) \right> 
   \nonumber \\
& = &-g^{\mu\nu}T_1 + v^{\mu}v^{\nu}T_2 
   + i \epsilon^{\mu\nu\alpha\beta}q_{\alpha}v_{\beta}T_3 
   + q^{\mu}q^{\nu}T_4 + (q^{\mu}v^{\nu}+v^{\mu}q^{\nu})T_5,
\end{eqnarray}
where $J^\mu$ is a $b \to f$ weak transition current.  The time
ordered product can be expressed as a sum over hadronic or partonic
intermediate states.  The sum over hadronic states includes the matrix
element $\left< H|J|B \right>$.  After inserting a complete set of
states and contracting with a four-vector pair $a_{\mu}^*a_{\nu}$, we
obtain:
\begin{eqnarray}\label{tmnexp}
T(\epsilon) &=& \frac1{2M_B}\sum_X (2\pi)^3\delta^3(\vec{p}_X+\vec q) 
  \frac{\left|\left<X\right|a\cdot J\left|B\right>\right|^2}{E_X-E_H-\epsilon}
  \nonumber\\ 
&& + \frac1{2M_B}\sum_X (2\pi)^3\delta^3(\vec{p}_X-\vec q) 
  \frac{\left|\left<B\right|a\cdot J\left|X\right>\right|^2}
  {\epsilon+E_X+E_H-2M_B},
\end{eqnarray}
where $T(\epsilon) \equiv a_{\mu}^*T^{\mu\nu}a_{\nu}$,
$\epsilon=M_B-E_H-v \cdot q$, and the sum over $X$ includes the usual
$\int d^3p/2E_X$ for each particle in the state $X$.  We choose to
work in the rest frame of the $B$ meson, $p = M_B v$, with the $z$
axis pointing in the direction of $\vec q$.  We hold $q_3$ fixed while
analytically continuing $v \cdot q$ to the complex plane.  $E_H =
\sqrt{M_H^2+q_3^2}$ is the $H$ meson energy.  There are two cuts in
the complex $\epsilon$ plane, $0<\epsilon<\infty$, corresponding to
the decay process $b \to f$, and $-\infty<\epsilon<-2E_H$,
corresponding to two $b$ quarks and a $\bar f$ quark in the final
state.  The second cut will not be important for our discussion.

The integral over $\epsilon$ of the time ordered product,
$T(\epsilon)$, times a weight function, $\epsilon^n
W_\Delta(\epsilon)$, can be computed perturbatively in QCD
\cite{BR9798,BLRW97}.  For simplicity, we pick the weight function
$W_\Delta(\epsilon) = \theta(\Delta-\epsilon)$, which corresponds to
summing over all hadronic resonances up to the excitation energy
$\Delta$ with equal weight.  Relating the integral with the hard
cutoff to the exclusive states requires local duality at the scale
$\Delta$.  Therefore, $\Delta$ must be chosen large enough so that the
structure functions can be calculated perturbatively.

Taking the zeroth moment of $T(\epsilon)$, we get
\begin{eqnarray}
\label{zerothmmnt}
M_0 
&\equiv& 
\frac1{2\pi i}\int_{C}d\epsilon\,\theta(\Delta-\epsilon)\,T(\epsilon) 
  \nonumber \\
&=&
\frac{\left | \langle X| a\cdot J | B\rangle \right |^2}{4 M_B E_H}
+
{\sum_{X \not= H}}^{\prime} \theta(E_X - E_H - \Delta) 
  (2 \pi)^3 \delta^3(\vec{q} + \vec{p}_X)
  \frac{\left | \langle X| a\cdot J | B\rangle \right |^2}{2 M_B},
  \nonumber
\end{eqnarray}
where the primed summation means a sum over all the kinematically
allowed states except the $H$ meson.  So,
\begin{eqnarray}
\label{factor}
\frac{\left | \langle X| a\cdot J | B\rangle \right |^2}{4 M_B E_H
  \epsilon} &=&
M_0 - {\sum_{X \not= H}}^{\prime} \theta(E_X - E_H - \Delta) 
  (2 \pi)^3 \delta^3(\vec{q} + \vec{p}_X)
  \frac{\left | \langle X| a\cdot J | B\rangle \right |^2}{2 M_B}.
\end{eqnarray}

On the other hand, the first moment of $T(\epsilon)$ gives
\begin{eqnarray}
\label{firstmmnt}
M_1 
&\equiv&
  \frac1{2\pi i}\int_C d\epsilon\,\epsilon\,\theta(\Delta-\epsilon)T(\epsilon)
  \nonumber\\
&=&
  {\sum_{X\not=H}}^{\prime} \theta(\Delta-E_X+E_H)\,(E_X-E_H)
    \,(2\pi)^3 \delta^3(\vec{q} + \vec{p}_X)
  \frac{\left|\left<X\right|a\cdot J\left|B\right>\right|^2}{4M_B E_X}
  \nonumber \\
&& \! \! \! \! \! \! \! \! \! \! \! \! 
\left\{
\begin{array}{ll}
\leq
  (E_{max}-E_H) \displaystyle{\sum_{X\not=H}}^{\prime} \theta(\Delta-E_X+E_H)
    (2\pi)^3 \delta^3(\vec{q} + \vec{p}_X)
  \displaystyle\frac{\left|\left<X\right|a \cdot J\left|
    B\right>\right|^2}{4M_B E_X}, \\
\geq
  (E_1-E_H) \displaystyle{\sum_{X\not=H}}^{\prime} \theta(\Delta-E_X+E_H)
    (2\pi)^3 \delta^3(\vec{q} + \vec{p}_X)
  \displaystyle\frac{\left|\left<X\right|a \cdot J\left|
    B\right>\right|^2}{4M_B E_X}.
\end{array}
\right.
\end{eqnarray}
where $E_{max}$ and $E_1$ denote the highest energy state
kinematically allowed and the first excited state that is more massive
than $H$ meson, respectively.  Here the validity of the second
inequality relies on the assumption that multiparticle final states
with energy less than $E_1$ contribute negligibly.  This assumption is
true in large $N_c$, and is also confirmed by current experimental
data.  However, the first inequality is valid without any further
assumption.

From Eq.~(\ref{factor}) and the first inequality in
Eq.~(\ref{firstmmnt}), one can get an upper bound on the matrix
element $\left| \left< H \right| a \cdot J \left| B \right> \right|^2
/ 4M_B E_H$,
\begin{equation}
\label{upbound}
\frac{\left| \left< H \right| a \cdot J \left| B \right> 
   \right|^2}{4M_B E_H}
   \leq \frac{1}{2\pi i} \int_{C} d\epsilon \, \theta(\Delta-\epsilon) \, 
   T(\epsilon) \left( 1-\frac{\epsilon}{E_{max}-E_H} 
   \right). \nonumber
\end{equation}
Dropping $\epsilon / (E_{max}-E_H)$ on the right hand side gives the
previously derived upper bound \cite{BSUV95,BR9798,BLRW97}.  Since
$E_{max}-E_H$ is of order $m_Q$ and the first moment, $M_1$, is of
order $1/m_Q$ and positive definite, this extra term makes the new
upper bound smaller than the old one at order $1/m_Q^2$.  Perturbative
corrections will also modify the new bound at order $\alpha_s/m_Q$.

Similarly, a lower bound can be formed by combining Eq.~(\ref{factor}) and the
second inequality in Eq.~(\ref{firstmmnt}) to be
\begin{equation}
\label{lowbound}
\frac{\left| \left< H \right| a \cdot J \left| B \right> 
   \right|^2}{4M_B E_H}
   \geq \frac{1}{2\pi i} \int_{C} d\epsilon \, \theta(\Delta-\epsilon) \, 
   T(\epsilon) \left( 1-\frac{\epsilon}{E_1-E_H} 
   \right). \nonumber
\end{equation}
Therefore, we find the bounds
\begin{eqnarray}
\label{uplowbound}
\frac{1}{2\pi i} \int_{C} d\epsilon \, \theta(\Delta-\epsilon) \, T(\epsilon)
   \left( 1-\frac{\epsilon}{E_1-E_H} \right)
&\leq& \frac{\left| \left< H(v') \right| a \cdot J \left| B(v) \right>
   \right|^2}{4M_B E_H} \nonumber \\
&\leq& \; \frac{1}{2\pi i} \int_{C} d\epsilon \, \theta(\Delta-\epsilon) \, 
   T(\epsilon) \left( 1-\frac{\epsilon}{E_{max}-E_H} 
   \right).
\end{eqnarray}
Since $1/(E_1-E_H) \sim 1/\Lambda_{\rm QCD}$, the lower bounds will be
good to one less order in $1/m_Q$ than the upper bound.

As emphasized in \cite{BLRW97}, the old upper bound is essentially
model independent while the lower bound relies on the assumption about
the final state spectrum.  The new upper bound provided here is also
model independent.  These bounds are valid for both heavy mesons and
baryons.  (For baryons, a spin sum
$\frac{M_H}{2j+1}\sum_{S,S^{\prime}}$ needs to be included in front of
the bounded factor.)


Great interest has been paid to the semileptonic exclusive decay rate
of $B \to D^* l \bar\nu$ from which $|V_{cb}|$ can be extracted
\cite{N91}.  As an example, we now focus on the case that $H$ is the
$D^*$ meson and give, in particular, the upper bounds on the form
factors $h_{A_1}$ and $h_V$.  The hadronic matrix element for the
semileptonic decay of a $B$ meson into a vector meson $D^*$ may be
parameterized as
\begin{eqnarray}
\label{param}
\frac{\left< D^*(v^\prime,\varepsilon) \mid V^\mu-A^\mu \mid B(p)\right>}
  {\sqrt{M_{D^*} M_B}} 
&=& 
- h_{A_1}(\omega) \, (\omega +1) \varepsilon^{*\mu}
+ \left[ h_{A_2}(\omega) v^\mu + h_{A_3}(\omega) v^{\prime\mu} \right]
     v \cdot \varepsilon^* \nonumber \\
&&
+ i h_V(\omega) \epsilon^{\mu\nu\alpha\beta}
   \varepsilon^*_\nu v^\prime_\alpha v_\beta,
\end{eqnarray}
where $v'$ is the velocity of the final state meson, and the variable
$\omega = v \cdot v^{\prime}$ is a measure of the recoil.  One may
relate $\omega$ to the momentum transfer $q^2$ by $\omega =
(M_B^2+{M_{D^*}}^2-q^2)/(2M_B M_{D^*})$.  Therefore, with a proper
choice of the current $J^{\mu}$ and the four vector $a^{\mu}$, one may
readily single out the form factors, $h_{A_1}$ and $h_V$, and
establish corresponding bounds, as was done in references
\cite{BR9798,BLRW97,CL99}.  Nonperturbative corrections to the
structure functions can be found in references
\cite{BKSV94,MW94,BGM96}, whereas complete ${\mathcal O}(\alpha_s)$
corrections are given in references \cite{BLRW97,CL99}.

To obtain the bounding curves within the kinematic range, $1 < \omega
\lesssim 1.25$, we will expand in $\alpha_s$, $\Lambda_{\rm QCD}/m_Q$
and $\omega-1$.  For both the upper and lower bounds, we will keep
perturbative corrections up to order $\alpha_s (\omega-1)$, but drop
terms of order $\alpha_s (\omega-1)^2$, $\alpha_s^2$, and $\alpha_s
\Lambda_{\rm QCD}/m_Q$.  We will calculate to order $1/m_Q^2$ for the
upper bounds, but only to order $1/m_Q$ for the lower bounds.

Both the old and new upper bounds along with the lower bound on
$h_{A_1}$ are shown \footnote{For the figures we take $m_b = 4.8 {\rm\
GeV}$, $m_c = 1.4 {\rm\ GeV}$, $\alpha_s = 0.3$ (corresponding to a
scale of about $2 {\rm\ GeV}$), $\bar\Lambda=0.4 {\rm\ GeV}$,
$\lambda_1=-0.2 {\rm\ GeV^2}$, $\lambda_2 = 0.12 {\rm\ GeV^2}$ and
$\Delta = 1{\rm\ GeV}$.} in Fig.~1.  In this and the next example, the
corresponding first excited state more massive than $D^*$ that
contributes to the sum rule is the $J^P = 1^+$ state, {\it i.e.}, the
$D_1$ meson, and $E_{max}$ is taken to be $M_B$ in the limit of no
energy transfer to the leptonic sector.
\begin{figure}[t]
\centerline{\epsfysize=11truecm  \epsfbox{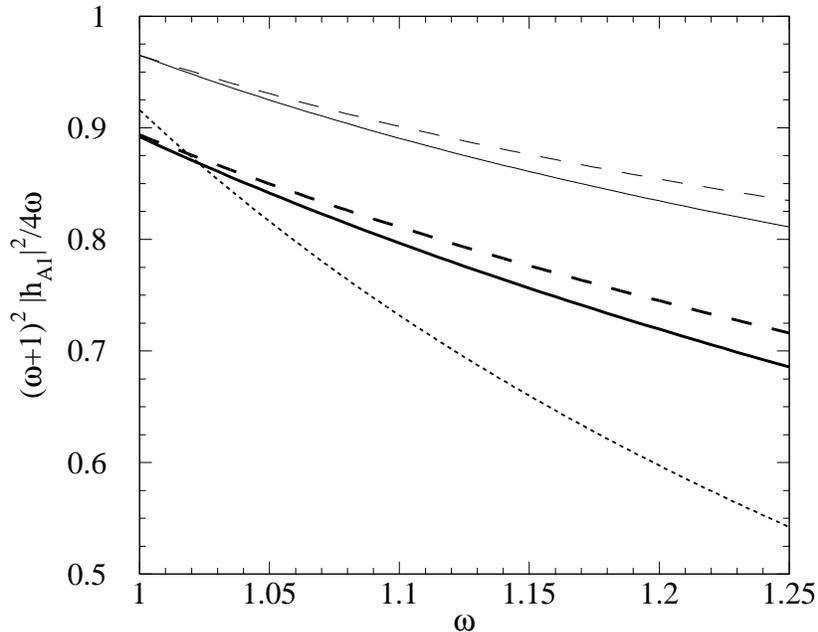} }
\tighten{
\caption[]{\it The upper bound on $(\omega+1)^2 \left| h_{A_1}(\omega)
\right|^2/(4\omega)$.  The thick solid (dashed) curve is the new (old)
upper bound to ${\mathcal O}(1/m_Q^2)$ including perturbative
corrections.  The thin solid (dashed) curve is the upper bound to
${\mathcal O}(1/m_Q^2)$ without perturbative corrections. The dotted
line is the lower bound to ${\mathcal O}(1/m_Q)$ including
perturbative corrections.}}
\end{figure}
The upper and lower bounds for $(\omega^2-1) \left| h_V(\omega)
\right|^2/(4\omega)$ are shown in Fig.~2.  
\begin{figure}[t]
\centerline{\epsfysize=11truecm  \epsfbox{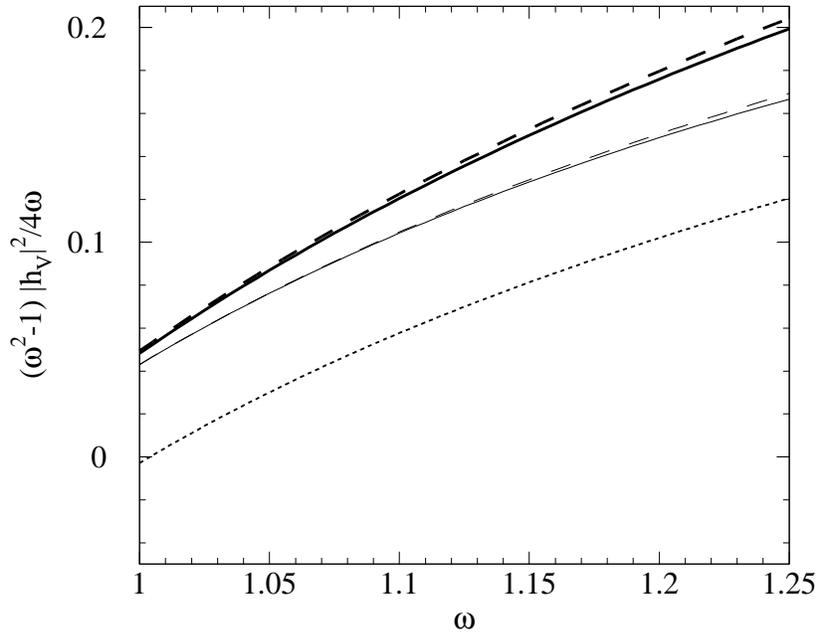} }
\tighten{
\caption[]{\it The upper bound on $(\omega^2-1) \left| h_V(\omega)
\right|^2/(4\omega)$.  The thick solid (dotted) curve is the new (old)
upper bound to ${\mathcal O}(1/m_Q^2)$ including perturbative
corrections.  The thin solid (dashed) curve is the upper bound to
${\mathcal O}(1/m_Q^2)$ without perturbative corrections.  The dotted
line is the lower bound to ${\mathcal O}(1/m_Q)$ including
perturbative corrections.}}
\end{figure}

In both diagrams, the thick solid (dashed) curve is the new (old)
upper bound including perturbative corrections.  The thin solid
(dashed) curve is the upper bound without perturbative corrections.
At large recoil, the new bound improves the upper limit by more than
$4\%$ in Fig.~1 and by about $3\%$ in Fig.~2.


This work provides tighter upper bounds on weak decay form factors.
The new upper bounds are compared with the old ones on, in particular,
the $B \to D^*$ form factors, $h_{A_1}$ and $h_V$.  Their difference
is due to the $1/m_Q^2$ nonperturbative corrections and $\alpha_s$
corrections that are suppressed by $1/M_Q$.  The difference of higher
order $1/m_Q$ corrections between the old and new bounds will be more
significant.


\acknowledgments

The author would like to thank Fred Gilman, Ira Rothstein and Adam
Leibovich for useful comments and discussions.  This work was
supported in part by the Department of Energy under Grant
No. DE-FG02-91ER40682.


{\tighten

}

\end{document}